\documentclass[11pt,a4paper]{article}
\usepackage{amsfonts}
\usepackage{amsmath}
\usepackage{a4wide}
\usepackage{amssymb}
\usepackage{graphicx}

\newtheorem{theorem}{Theorem}

\newtheorem{corollary}{Corollary}

\newtheorem{definition}{Definition}

\newtheorem{proposition}{Proposition}
\newtheorem{remark}{Remark}

\makeatletter
\def\@removefromreset#1#2{\let\@tempb\@elt
     \def\@tempa#1{@&#1}\expandafter\let\csname @*#1*\endcsname\@tempa
     \def\@elt##1{\expandafter\ifx\csname @*##1*\endcsname\@tempa\else
    \noexpand\@elt{##1}\fi}     \expandafter\edef\csname cl@#2\endcsname{\csname cl@#2\endcsname}     \let\@elt\@tempb
     \expandafter\let\csname @*#1*\endcsname\@undefined}

\@removefromreset{equation}{section}

\@removefromreset{theorem}{section}
\makeatother
\input{tcilatex}
\begin{document}

\title{Nonsignaling as the consistency condition for local quasi classical
probability modelling of a general multipartite correlation scenario}
\author{Elena R. Loubenets \\
Applied Mathematics Department, Moscow State Institute \\
of Electronics and Mathematics, Moscow 109028, Russia}
\maketitle

\begin{abstract}
We specify for a general correlation scenario a particular type of a local
quasi hidden variable (LqHV) model [\emph{J. Math. Phys. }\textbf{53 }%
(2012), 022201] -- a deterministic LqHV model, where all joint probability
distributions of a correlation scenario are simulated via a single measure
space with a normalized bounded real-valued measure not necessarily positive
and random variables, each depending only on a setting of the corresponding
measurement at the corresponding site. We prove that an arbitrary
multipartite correlation scenario admits a deterministic LqHV model if and
only if all its joint probability distributions satisfy the consistency
condition constituting the general nonsignaling condition formulated in [%
\emph{J. Phys. A: Math. Theor.} \textbf{41 }(2008), 445303]. This
mathematical result specifies a new probability model that has the
measure-theoretic structure resembling the structure of the classical
probability model but incorporates the latter only as a particular case. The
local version of this \emph{quasi classical probability model} covers the
probabilistic description of every nonsignaling correlation scenario, in
particular, each correlation scenario on an multipartite quantum state.
\end{abstract}

\section{Introduction}

A possibility of the description of quantum measurements in terms of the
classical probability model has been a point of intensive discussions ever
since the seminal publications of von Neumann \cite{neumann}, Kolmogorov 
\cite{kolmogorov}, Einstein, Podolsky and Rosen (EPR) \cite{EPR} and Bell 
\cite{bell, bell1}.

Though, in the quantum physics literature, one can still find the misleading%
\footnote{%
On the misleading character of such statements, see also Ref. \cite%
{loubkhren}.} claims on a peculiarity of \textquotedblright quantum
probabilities" and \textquotedblright quantum events", the probabilistic
description of every quantum measurement satisfies the Kolmogorov axioms 
\cite{kolmogorov} for the theory of probability.

Namely, every measurement on a quantum system represented initially by a
state $\rho $ on a complex separable Hilbert space $\mathcal{H}$ is
described by the probability space\footnote{%
In the measure theory, this triple is called a measure space.} $(\Lambda ,%
\mathcal{F}_{\Lambda },\mathrm{tr}[\rho \mathrm{M}(\cdot )]),$ where $%
\Lambda $ is a set of measurement outcomes, $\mathcal{F}_{\Lambda }$ is a $%
\sigma $-algebra of observed events $F\subseteq \Lambda $ and $\mathrm{tr}%
[\rho \mathrm{M}(\cdot )]:\mathcal{F}_{\Lambda }\rightarrow \lbrack 0,1]$ is
the probability measure with values $\mathrm{tr}[\rho \mathrm{M}(F)],$ $F\in 
\mathcal{F}_{\Lambda },$ each defining the probability that, under this
quantum measurement, an outcome $\lambda $ belongs to a set $F\in \mathcal{F}%
_{\Lambda }.$ Here, \textrm{M }is a normalized ($\mathrm{M}(\Lambda )=%
\mathbb{I}_{\mathcal{H}})$ measure with values $\mathrm{M}(F),$ $F\in 
\mathcal{F}_{\Lambda },$ that are positive operators on $\mathcal{H}$ $%
\mathrm{-}$ that is, a normalized positive operator-valued (POV) measure%
\footnote{%
The description of a quantum measurement via a POV measure was introduced by
Davies and Lewis \cite{daviesLewis, davies}} on $(\Lambda ,\mathcal{F}%
_{\Lambda })$.

The measure-theoretic structure of the Kolmogorov axioms \cite{kolmogorov}
is crucial and the probabilistic description of each measurement in every
application field satisfies these probability axioms.

However, the classical probability model, which is also often named\footnote{%
In the probability theory, the term "Kolmogorov probability model" refers to
the probabilistic description of a measurement via the Kolmogorov axioms,
see, for example, in Ref. \cite{shiryaev}.} after Kolmogorov in the
mathematical physics literature and where system observables and states are
represented by random variables and probability measures on a single
measurable space $(\Omega ,\mathcal{F}_{\Omega }),$ describes correctly
randomness in the classical statistical mechanics and many other application
fields, \emph{but} \emph{fails} either to reproduce noncontextually \cite%
{kochen} the statistical properties of all quantum observables on a Hilbert
space of a dimension$\ \dim \mathcal{H}\geq 3$ or to simulate via random
variables, each depending only on a setting of the corresponding measurement
at the corresponding site, the probabilistic description of a quantum
correlation scenario on an arbitrary $N$-partite quantum state. For details
and references, see section 1.4 in \cite{holevo} and the introduction in 
\cite{loubenets3}.

The probabilistic description\footnote{%
For the general framework on the probabilistic description of multipartite
correlation scenarios, see Ref. \cite{loubenets3}.} of an arbitrary
multipartite correlation scenario cannot be also reproduced via the
classical probability model.

Note that, in the quantum theory literature, the interpretation of quantum
measurements in the classical probability terms is generally referred to as
a hidden variable (HV) model.

In \cite{10}, we have introduced for a general correlation scenario the
notion of \emph{a} \emph{local quasi hidden variable (LqHV) model,} where
locality and the measure-theoretic structure inherent to a local hidden
variable (LHV) model are preserved but positivity of a simulation measure is
dropped. We have proved \cite{10} that every quantum $S_{1}\times \cdots
\times S_{N}$-setting correlation scenario admits LqHV modelling and
specified the state parameter determining quantitatively a possibility of an 
$S_{1}\times \cdots \times S_{N}$-setting LHV description of an $N$-partite
quantum state.

In the present article, we develop further the LqHV approach introduced in 
\cite{10}. The paper is organized as follows.

In section 2, we specify for a general multipartite correlation scenario the
notion of a deterministic LqHV model\emph{, }where\emph{\ }all joint
probability distributions of a correlation scenario are simulated via a
single measure space with a normalized bounded real-valued measure and
random variables, each depending only on a setting of the corresponding
measurement at the corresponding site. We show that the existence for a
general correlation scenario of some LqHV model implies the existence for
this scenario of a deterministic LqHV model.

In section 3, we prove that an arbitrary multipartite correlation scenario
admits a deterministic LqHV model if an only if all its joint probability
distributions satisfy the consistency condition constituting the general
nonsignaling condition formulated by Eq. (10) in \cite{loubenets3}.

In section 4, we summarize the main mathematical results of the present
article and discuss their conceptual implication.

\section{A deterministic LqHV model}

Consider an $N$-partite correlation scenario, where each $n$-th of $N\geq 2$
parties (players) performs $S_{n}\geq 1$ measurements with outcomes $\lambda
_{n}\in \Lambda _{n}\ $of an arbitrary type and $\mathcal{F}_{\Lambda _{n}}$
is a $\sigma $-algebra of events $F_{n}\subseteq $ $\Lambda _{n}$ observed
at $n$-th site. We label each measurement at $n$-th site by a positive
integer $s_{n}=1,...,S_{n}$ and each of $N$-partite joint measurements,
induced by this correlation scenario and with outcomes $(\lambda _{1},\ldots
,\lambda _{N})\in \Lambda _{1}\times \cdots \times \Lambda _{N}$ -- by an $N$%
-tuple $(s_{1},...,s_{N}),$ where $n$-th component refers to a measurement
at $n$-th site.

For concreteness, we further specify an $S_{1}\times \cdots \times S_{N}$%
-setting correlation scenario by the symbol $\mathcal{E}_{S},$ where $%
S:=S_{1}\times \cdots \times S_{N}$, and denote by $P_{(s_{1},...,s_{N})}^{(%
\mathcal{E}_{S})}$ a probability measure, defined on the direct product%
\footnote{%
Recall \cite{dunford} that the product $\sigma $-algebra $\mathcal{F}%
_{\Lambda _{1}}\otimes \cdots \otimes \mathcal{F}_{\Lambda _{N}}$ on $%
\Lambda _{1}\times \cdots \times \Lambda _{N}$ is the smallest $\sigma $%
-algebra generated by the set of all rectangles $F_{1}\times \cdots \times
F_{N}\subseteq \Lambda _{1}\times \cdots \times \Lambda _{N}$ with
measurable "sides" $F_{n}\in \mathcal{F}_{\Lambda _{n}},$ $n=1,...,N.$} $%
(\Lambda _{1}\times \cdots \times \Lambda _{N},$ $\mathcal{F}_{\Lambda
_{1}}\otimes \cdots \otimes \mathcal{F}_{\Lambda _{N}})$ of measurable
spaces $(\Lambda _{n},\mathcal{F}_{\Lambda _{n}}),$ $n=1,...,N,$ and
describing an $N$-partite joint measurement $(s_{1},...,s_{N})$ under a
scenario $\mathcal{E}_{S}$.

\begin{remark}
The superscript $\mathcal{E}_{S}$ at notation $P_{(s_{1},...,s_{N})}^{(%
\mathcal{E}_{S})}$ indicates that, in contrast to a correlation scenario
represented by the so-called "nonsignaling boxes" \cite{11, 12} and
described by joint probability distributions $P_{(s_{1},...,s_{N})}^{(%
\mathcal{E}_{S})}\equiv P_{(s_{1},...,s_{N})}.$ $%
s_{1}=1,...,S_{1},...,s_{N}=1,...,S_{N},$ each depending only on settings of
the corresponding measurements at the corresponding sites, for a general
correlation scenario $\mathcal{E}_{S},$ each distribution $%
P_{(s_{1},...,s_{N})}^{(\mathcal{E}_{S})}$ may also depend on settings of
all (or some) other measurements. The latter is, for example, the case under
a classical correlation scenario with \textquotedblleft one-sided" or
"two-sided" memory \cite{13}.
\end{remark}

If, under an $N$-partite joint measurement $(s_{1},...,s_{N})$ of scenario $%
\mathcal{E}_{S}$ only outcomes of $M<N$ parties $1\leq n_{1}<...<n_{M}\leq N$
are taken into account while outcomes of all other parties are ignored, then
the joint probability distribution of outcomes observed at these $M$ sites
is described by the marginal probability distribution\ 
\begin{equation}
P_{(s_{1},...,s_{N})}^{(\mathcal{E}_{S})}(\Lambda _{1}\times \cdots \times
\Lambda _{n_{1}-1}\times \mathrm{d}\lambda _{_{n_{1}}}\times \Lambda
_{n_{1}+1}\times \cdots \times \Lambda _{n_{_{M}}-1}\times \mathrm{d}\lambda
_{_{n_{_{_{M}}}}}\times \Lambda _{n_{_{M}}+1}\times \cdots \times \Lambda
_{_{N}}).  \label{1}
\end{equation}%
In particular, 
\begin{equation}
P_{(s_{1},...,s_{N})}^{(\mathcal{E}_{S})}(\Lambda _{1}\times \cdots \times
\Lambda _{n-1}\times \mathrm{d}\lambda _{n}\times \Lambda _{n+1}\times
\cdots \times \Lambda _{_{N}})  \label{2}
\end{equation}%
is the probability distribution of outcomes observed at $n$-th site under a
joint measurement $(s_{1},...,s_{N})$ of scenario $\mathcal{E}_{S}.$

\begin{remark}
Throughout this paper, for a measure $\tau $ on the direct product $(\Lambda
\times \cdots \times \Lambda ^{\prime }$, $\mathcal{F}_{\Lambda }\otimes
\cdots \otimes \mathcal{F}_{\Lambda ^{\prime }})$ of some measurable spaces,
we often use notation $\tau (\mathrm{d}\lambda \times \cdots \times \mathrm{d%
}\lambda ^{\prime })$ outside of an integral. This allows us to specify
easily the structure of different marginals of $\tau $.
\end{remark}

For the probabilistic description of a general correlation scenario,
consider the following simulation model introduced in \cite{10}.

\begin{definition}
\cite{10} An $S_{1}\times ...\times S_{N}$-setting correlation scenario $%
\mathcal{E}_{S},$ with joint probability distributions $%
P_{(s_{1},...,s_{N})}^{(\mathcal{E}_{S})},$ \ $%
s_{1}=1,...,S_{1},...,s_{N}=1,...,S_{N},$ and outcomes $(\lambda _{1},\ldots
,\lambda _{N})\in \Lambda _{1}\times \cdots \times \Lambda _{N}$ of an
arbitrary type admits a local quasi hidden variable (LqHV) model if all of
its joint probability distributions admit the representation 
\begin{align}
P_{(s_{1},...,s_{N})}^{(\mathcal{E}_{S})}(F_{1}\times \cdots \times F_{N})&
=\dint\limits_{\Omega }P_{1}^{(s_{_{1}})}(F_{1}\text{ }|\text{ }\omega
)\cdot \ldots \cdot P_{N}^{(s_{_{N}})}(F_{N}\text{ }|\text{ }\omega )\text{ }%
\nu _{\mathcal{E}_{S}}(\mathrm{d}\omega ),  \label{3} \\
F_{1}& \in \mathcal{F}_{\Lambda _{1}},...,F_{N}\in \mathcal{F}_{\Lambda
_{N}},  \notag
\end{align}%
in terms of a single measure space $\left( \Omega ,\mathcal{F}_{\Omega },\nu
_{\mathcal{E}_{S}}\right) $ with a normalized bounded real-valued measure $%
\nu _{\mathcal{E}_{S}}$ and conditional probability measures $%
P_{n}^{(s_{_{n}})}(\cdot $ $|$ $\omega ):\mathcal{F}_{\Lambda
_{n}}\rightarrow \lbrack 0,1],$ defined $\nu _{_{\mathcal{E}_{S}}}$-a. e.
(almost everywhere) on $\Omega $ and such that, for each $s_{n}=1,...,S_{n}$
and every $n=1,...,N,$ the function $P_{n}^{(s_{_{n}})}(F_{n}$ $|$ $\cdot
):\Omega \rightarrow \lbrack 0,1]$ is measurable for each $F_{n}\in \mathcal{%
F}_{\Lambda _{n}}.$\medskip
\end{definition}

In a triple $\left( \Omega ,\mathcal{F}_{\Omega },\nu \right) $ representing
a measure space, $\Omega $ is a non-empty set, $\mathcal{F}_{\Omega }$ is a $%
\sigma $-algebra of subsets of $\Omega $ and $\nu $ is a measure on a
measurable space $\left( \Omega ,\mathcal{F}_{\Omega }\right) .$ A
real-valued measure $\nu $ is called normalized if $\nu (\Omega )=1$ and
bounded \cite{dunford} if $\left\vert \nu (F)\right\vert \leq M<\infty $ for
all $F\in \mathcal{F}_{\Omega }.$ Note that each bounded real-valued measure 
$\nu $ admits \cite{dunford} the Jordan decomposition $\nu =\nu ^{+}-\nu
^{=} $ via positive measures%
\begin{equation}
\nu ^{+}(F):=\sup_{F^{\prime }\in \mathcal{F}_{\Omega },F^{\prime }\subseteq
F}\nu (F^{\prime }),\text{ \ \ \ }\nu ^{-}(F):=-\inf_{F^{\prime }\in 
\mathcal{F}_{\Omega },F^{\prime }\subseteq F}\nu (F^{\prime }),\text{ \ \ }%
\forall F\in \mathcal{F}_{\Omega },  \label{33}
\end{equation}%
with disjoint supports.

We stress that, in an LqHV model (\ref{3}), a normalized bounded real-valued
measure $\nu _{\mathcal{E}_{S}}$ has \emph{a simulation character }and may,
in general, depend (via the subscript $\mathcal{E}_{S}$) on measurement
settings at all (or some) sites, as an example, see measure (39) in \cite{10}%
.

The structure of each LqHV model is such that though some values of a
simulation measure $\nu _{\mathcal{E}_{S}}$ may be negative, the integral
standing in the right-hand side of representation (\ref{3})\ is \emph{%
non-negative} for all $F_{1}\in \mathcal{F}_{\Lambda _{1}},...,F_{N}\in 
\mathcal{F}_{\Lambda _{N}}.$

If, for a correlation scenario $\mathcal{E}_{S}$, there exists
representation (\ref{3}), where a normalized bounded real-valued measure $%
\nu _{\mathcal{E}_{S}}$ is positive (hence, is a probability measure), then
this scenario admits a local hidden variable (LHV) model\emph{\ }formulated%
\emph{\ }for a general case by Eq. (26) in \cite{loubenets3}.

As it is discussed in detail in \cite{10}, the concept of an LqHV model
incorporates as particular cases and generalizes in one whole both types of
simulation models known in the literature -- an LHV model and an affine
model \cite{kaplan}. Note that the latter model, where all distributions $%
P_{(s_{1},...,s_{N})}^{(\mathcal{E}_{S})}$ of a scenario $\mathcal{E}_{S}$
are expressed via the affine sum of some LHV\ distributions, is in principle
built up on the concept of an LHV model.

Introduce now the following special type of an LqHV model.

\begin{definition}
An LqHV model (\ref{3}) is called deterministic if there exist $\mathcal{F}%
_{\Omega }/\mathcal{F}_{\Lambda _{n}}$- measurable functions (random
variables) $f_{n,s_{n}}:$ $\Omega \rightarrow \Lambda _{n},$ such that, in
representation (\ref{3}), all conditional probability measures $%
P_{n}^{(s_{n})}(\cdot |\omega ),$ $s_{n}=1,...,S_{n},$ $n=1,...,N,$ have the
special form 
\begin{equation}
P_{n}^{(s_{n})}(F_{n}\text{ }|\text{ }\omega )=\chi
_{f_{n,s_{n}}^{-1}(F_{n})}(\omega ),\text{ \ \ }\forall F_{n}\in \mathcal{F}%
_{\Lambda _{n}},  \label{4'}
\end{equation}%
$\nu _{\mathcal{E}_{S}}$-a. e. on $\Omega .$
\end{definition}

Here, $f^{-1}(F)=\left\{ \omega \in \Omega \mid f(\omega )\in F\right\} $ is
the preimage of a set $F\in \mathcal{F}_{\Lambda }$ under a mapping $%
f:\Omega \rightarrow \Lambda $ and $\chi _{D}(\cdot )$ is the indicator
function of a subset $D\subseteq \Omega ,$ that is, $\chi _{D}(\omega )=1$
for $\omega \in D$ and $\chi _{D}(\omega )=0$ for $\omega \notin D.$

The notion of a deterministic LqHV model generalizes the concept of a
deterministic\footnote{%
The terms "deterministic HV model" and "stochastic HV model" were first
introduced by Fine \cite{fine} for a bipartite scenario with two settings
and two outcomes per site.} LHV model formulated for a general multipartite
correlation scenario in section 4 of \cite{loubenets3}.

From (\ref{3}) and (\ref{4'}) it follows that if an $S_{1}\times \cdots
\times S_{N}$-setting correlation scenario $\mathcal{E}_{S}$ admits a
deterministic LqHV model, then all its joint probability distributions $%
P_{(s_{1},...,s_{N})}^{(\mathcal{E}_{S})}$ admit the representation%
\begin{align}
P_{(s_{1},...,s_{N})}^{(\mathcal{E}_{S})}(F_{1}\times \cdots \times F_{N})&
=\dint\limits_{\Omega }\chi _{f_{1,s_{1}}^{-1}(F_{1})}(\omega )\cdot \ldots
\cdot \chi _{f_{N,s_{N}}^{-1}(F_{N})}(\omega )\text{ }\nu _{\mathcal{E}_{S}}(%
\mathrm{d}\omega )  \label{4} \\
& =\nu _{\mathcal{E}_{S}}\left( f_{1,s_{1}}^{-1}(F_{1})\cap \cdots \cap
f_{N,s_{_{N}}}^{-1}(F_{N})\right) ,  \notag \\
F_{1}& \in \mathcal{F}_{\Lambda _{1}},...,F_{N}\in \mathcal{F}_{\Lambda
_{N}},  \notag
\end{align}%
via a normalized bounded real-valued measure $\nu _{\mathcal{E}_{S}}$ on
some measurable space $(\Omega ,\mathcal{F}_{\Omega })$ and random variables 
$f_{n,s_{n}}:\Omega \rightarrow \Lambda _{n},$ each depending only on a
setting of $s_{n}$-th measurement at $n$-th site.

In a deterministic LqHV model, the relation between a simulation measure $%
\nu _{\mathcal{E}_{S}}$\ and random variables $f_{n,s_{n}},$ $%
s_{n}=1,...,S_{n},$ $n=1,...,N,$ modelling scenario measurements is such
that \emph{the joint probabilities of scenario events are reproduced due to (%
\ref{4}) only via non-negative values of }$\nu _{\mathcal{E}_{S}}$.

Representation (\ref{4}), in turn, implies that, for arbitrary bounded
measurable real-valued functions $\varphi _{n}:\Lambda _{n}\rightarrow 
\mathbb{R},$ $n=1,...,N,$ the product expectation 
\begin{equation}
\left\langle \varphi _{_{1}}(\lambda _{1})\cdot ...\cdot \varphi
_{_{N}}(\lambda _{N})\right\rangle _{(s_{1},...,s_{N})}^{(\mathcal{E}%
_{S})}:=\dint\limits_{\Lambda }\varphi _{_{1}}(\lambda _{1})\cdot ...\cdot
\varphi _{_{N}}(\lambda _{N})P_{(s_{1},...,s_{N})}^{(\mathcal{E}_{S})}(%
\mathrm{d}\lambda _{1}\times \cdots \times \mathrm{d}\lambda _{N})
\label{gg}
\end{equation}%
takes the form%
\begin{equation}
\left\langle \varphi _{_{1}}(\lambda _{1})\cdot ...\cdot \varphi
_{_{N}}(\lambda _{N})\right\rangle _{(s_{1},...,s_{N})}^{(\mathcal{E}%
_{S})}=\dint (\varphi _{_{1}}\circ f_{1,s_{1}})(\omega )\cdot ...\cdot
(\varphi _{_{N}}\circ f_{N,s_{N}})(\omega )\text{ }\nu _{\mathcal{E}_{S}}(%
\mathrm{d}\omega ),  \label{5}
\end{equation}%
which differs from the form of the product expectations in a deterministic
LHV model (see Eq. (31) in \cite{loubenets3}) only by the fact that a
normalized bounded real-valued measure $\nu _{\mathcal{E}_{S}}$ in (\ref{5})
does not need to be positive.

Recall \cite{loubenets3} that, for a given correlation scenario, a
deterministic LHV model constitutes the version of the local classical
probability model, where \emph{only} the observed joint probability
distributions are reproduced.

Therefore, a deterministic LqHV model (\ref{4}) corresponds to the local 
\emph{quasi classical probability model, }where, in contrast to the local
classical probability model, an "underlying" probability space is replaced\
by a measure space $(\Omega ,\mathcal{F}_{\Omega },\nu )$\ with a normalized
bounded real-valued measure $\nu $\ not necessarily positive and where:%
\newline
(i) observables with a value space $(\Lambda ,\mathcal{F}_{\Lambda })$ are
represented \emph{only} by such random variables $f:\Omega \rightarrow
\Lambda $ for which $\nu (f^{-1}(F))\geq 0,$ $\forall F\in \mathcal{F}%
_{\Lambda };$\newline
(ii) a joint measurement of two observables $f_{1},$ $f_{2},$ each with a
value spaces $(\Lambda _{n},\mathcal{F}_{\Lambda _{n}}),$ is possible\emph{\
if and only if} $\nu (f_{1}^{-1}(F_{1})\cap f_{2}^{-1}(F_{2}))\geq 0$ for
all $F_{n}\in \mathcal{F}_{\Lambda _{n}}.\medskip $

The following statement is proved in appendix A.

\begin{proposition}
If an $S_{1}\times ...\times S_{N}$-setting correlation scenario $\mathcal{E}%
_{S}$ admits some LqHV model (\ref{3}), then it also admits a deterministic
LqHV model (\ref{4}).\medskip
\end{proposition}

This statement and theorem 1 of Ref. \cite{10} imply.

\begin{proposition}
An $S_{1}\times ...\times S_{N}$-setting correlation scenario $\mathcal{E}%
_{S}$ admits a deterministic LqHV model (\ref{4}) if and only if, on the
direct product space $(\Lambda _{1}^{S_{1}}\times \cdots \times \Lambda
_{N}^{S_{N}},$ $\mathcal{F}_{\Lambda _{1}}^{\otimes S_{1}}\otimes \cdots
\otimes \mathcal{F}_{\Lambda _{N}}^{\otimes S_{N}})$, there exists a
normalized bounded real-valued measure\footnote{%
See remark 2.} 
\begin{align}
& \mu _{\mathcal{E}_{S}}\left( \mathrm{d}\lambda _{1}^{(1)}\times \cdots
\times \mathrm{d}\lambda _{1}^{(S_{1})}\times \cdots \times \mathrm{d}%
\lambda _{N}^{(1)}\times \cdots \times \mathrm{d}\lambda
_{N}^{(S_{N})}\right) ,  \label{11'} \\
\lambda _{n}^{(s_{n})}& \in \Lambda _{n},\text{ \ \ }s_{n}=1,...,S_{n},\text{
\ \ }n=1,...,N,  \notag
\end{align}%
returning all joint probability distributions $P_{(s_{1},...,s_{N})}^{(%
\mathcal{E}_{S})}$ of a scenario $\mathcal{E}_{S}$ as the corresponding
marginals.
\end{proposition}

\section{The general consistency theorem}

Let us now analyze, under what condition on joint probability distributions,
an arbitrary multipartite correlation scenario admits a deterministic LqHV
model.

Suppose that, under an $S_{1}\times \cdots \times S_{N}$-setting correlation
scenario $\mathcal{E}_{S},$ for all joint measurements $%
(s_{1},...,s_{_{N}}), $ $(s_{1}^{\prime },...,s_{_{N}}^{\prime })$ with $%
1\leq M<N$ common settings $s_{n_{1}},...,s_{n_{_{M}}}$ at arbitrary sites $%
1\leq n_{1}<\cdots <n_{M}\leq N,$ the marginal probability distributions (%
\ref{1}) of outcomes observed at these sites coincide, that is:%
\begin{align}
& P_{(s_{1},...,s_{N})}^{(\mathcal{E}_{S})}(\Lambda _{1}\times \cdots \times
\Lambda _{n_{1}-1}\times \mathrm{d}\lambda _{_{n_{1}}}\times \Lambda
_{n_{1}+1}\times \cdots \times \Lambda _{n_{_{M}}-1}\times \mathrm{d}\lambda
_{_{n_{_{_{M}}}}}\times \Lambda _{n_{_{M}}+1}\times \cdots \times \Lambda
_{_{N}})  \label{11} \\
&  \notag \\
& =P_{(s_{1}^{\prime },..,s_{_{N}}^{\prime })}^{(\mathcal{E}_{S})}(\Lambda
_{1}\times \cdots \times \Lambda _{n_{1}-1}\times \mathrm{d}\lambda
_{_{n_{1}}}\times \Lambda _{n_{1}+1}\times \cdots \times \Lambda
_{n_{_{M}}-1}\times \mathrm{d}\lambda _{_{n_{_{_{M}}}}}\times \Lambda
_{n_{_{M}}+1}\times \cdots \times \Lambda _{_{N}}).  \notag
\end{align}

As we discuss this in section 3 of \cite{loubenets3}, for a general
correlation scenario $\mathcal{E}_{S}$ with a finite number of measurement
settings at each site, condition (\ref{11}) does not automatically imply
that the coinciding marginals, standing in the left-hand side and the
right-hand side of Eq. (\ref{11}), depend only on settings of measurements $%
s_{n_{1}},...,s_{n_{_{M}}}$ at sites $1\leq n_{1}<\cdots $ $<n_{M}\leq N.$

This means that condition (\ref{11}) should be distinguished from the
condition 
\begin{align}
& P_{(s_{1},..,s_{_{N}})}^{(\mathcal{E}_{S})}(\Lambda _{1}\times \cdots
\times \Lambda _{n_{1}-1}\times \mathrm{d}\lambda _{_{n_{1}}}\times \Lambda
_{n_{1}+1}\times \cdots \times \Lambda _{n_{_{M}}-1}\times \mathrm{d}\lambda
_{_{n_{_{_{M}}}}}\times \Lambda _{n_{_{M}}+1}\times \cdots \times \Lambda
_{_{N}})  \label{12} \\
& \equiv P_{(s_{n_{1}},...,s_{n_{_{M}}})}(\mathrm{d}\lambda
_{n_{_{1}}}\times \cdots \times \mathrm{d}\lambda _{n_{_{M}}}),  \notag \\
s_{1}& =1,...,S_{1},...,s_{N}=1,...,S_{N},\text{ \ \ }M=1,...,N,  \notag
\end{align}%
usually argued in the literature to follow if $M<N$ from condition (\ref{11}%
).

Though condition (\ref{12}) implies condition (\ref{11}), the converse is
not, in general, true, see proposition 1 in Ref. \cite{loubenets3}.

In view of their physical interpretations discussed in detail in \cite%
{loubenets3}, we call conditions (\ref{11}), (\ref{12}) as the nonsignaling
condition and the EPR locality condition, respectively. Moreover, since, in
the literature on quantum information\footnote{%
See, for example, Refs. \cite{11, 12, kaplan} and therein.}, specifically
the joint combination of conditions (\ref{11}), (\ref{12}) is often called
as nonsignaling, in order to exclude a possible misunderstanding, we further
refer to the consistency condition (\ref{11}) as \emph{the} \emph{general
nonsignaling condition.}

We stress -- \emph{the} \emph{nonsignaling condition in the sense of Refs. 
\cite{11, 12} implies the general nonsignaling condition (\ref{11}), but the
converse of this statement is not, in general, true. }

The following theorem is proved in appendix B.

\begin{theorem}
An $S_{1}\times \cdots \times S_{N}$-setting correlation scenario $\mathcal{E%
}_{S}$ admits a deterministic LqHV model (\ref{4}) if and only if all its
joint probability distributions $P_{(s_{1},...,s_{N})}^{(\mathcal{E}_{S})},$ 
$s_{1}=1,...,S_{1},...,s_{N}=1,...,S_{N},$ satisfy the consistency condition
(\ref{11}) constituting the general nonsignaling condition formulated in
Ref. \cite{loubenets3}.\medskip
\end{theorem}

Consider, in particular, an $S_{1}\times \cdots \times S_{N}$-setting
correlation scenario performed on an $N$-partite quantum state $\rho $ on a
complex separable Hilbert space $\mathcal{H}_{1}\otimes \cdots \otimes 
\mathcal{H}_{N}$ and described by the joint probability distributions 
\begin{align}
& \mathrm{tr}[\rho \{\mathrm{M}_{1}^{(s_{1})}(F_{1})\otimes \cdots \otimes 
\mathrm{M}_{N}^{(s_{_{N}})}(F_{N})\}],  \label{17} \\
F_{n}& \in \mathcal{F}_{n},\text{ \ \ }s_{n}=1,...,S_{n},\text{ \ \ }%
n=1,...,N,  \notag
\end{align}%
where $\mathrm{M}_{n}^{(s_{n})}$ is a POV\footnote{%
For this notion, see the introduction.} measure on $(\Lambda _{n},\mathcal{F}%
_{\Lambda _{n}})$ representing on a Hilbert space $\mathcal{H}_{n}$ a
quantum measurement $s_{n}$ at $n$-th site.

Since every quantum correlation scenario (\ref{17}) satisfies condition (\ref%
{11}) (as well as condition (\ref{12})), theorem 1 implies.

\begin{corollary}
For every quantum state $\rho $ on a complex separable Hilbert space $%
\mathcal{H}_{1}\otimes \cdots \otimes \mathcal{H}_{N}$ and arbitrary
positive integers $S_{1},...,S_{N}\geq 1,$ the probabilistic description of
each quantum $S_{1}\times \cdots \times S_{N}$ -setting correlation scenario
(\ref{17}) admits a deterministic LqHV model.
\end{corollary}

In view of the above proposition 1, the statement of corollary 1 agrees with
the statement of theorem 2 in \cite{10}.

\section{Conclusions}

In the present paper, we have introduced (definition 2) the notion of a
deterministic LqHV model, where all joint probability distributions of a
multipartite correlation scenario are simulated via a \emph{single} \emph{%
measure space} $(\Omega ,\mathcal{F}_{\Omega },\nu ),$ with a normalized
bounded real-valued measure\ $\nu $ not necessarily positive, and random
variables which are \emph{local} in the sense that each of these random
variables depends only on a setting of the corresponding measurement at the
corresponding site.

We have proved (theorem 1) that a general $S_{1}\times \cdots \times S_{N}$%
-setting correlation scenario admits a deterministic LqHV model if and only
if all its joint probability distributions satisfy the consistency condition
(\ref{11}) constituting the general nonsignaling condition formulated in
Ref. \cite{loubenets3}.

This general result, in particular, implies (corollary 1) that the
probabilistic description of every $S_{1}\times \cdots \times S_{N}$-setting
correlation scenario (\ref{17}) on an $N$-partite quantum state admits
modelling in local quasi classical terms.

From the conceptual point of view, these mathematical results specify a new
probability model that has the measure-theoretic structure $(\Omega ,%
\mathcal{F}_{\Omega },\nu )$ resembling the structure of the classical
probability model but reduces to the latter iff a normalized bounded
real-valued measure $\nu $ is positive. In the frame of this \emph{quasi
classical probability model}: \newline
(i) observables with a value space $(\Lambda ,\mathcal{F}_{\Lambda })\ $are
represented \emph{only} by such random variables $f:\Omega \rightarrow
\Lambda $ for which $\nu (f^{-1}(F))\geq 0,$ $\forall F\in \mathcal{F}%
_{\Lambda };$ \newline
(ii) a joint measurement of two observables $f_{1},$ $f_{2},$ each with a
value space $(\Lambda _{n},\mathcal{F}_{\Lambda _{n}}),$ is possible \emph{%
if and only if }$\nu (f_{1}^{-1}(F_{1})\cap f_{2}^{-1}(F_{2}))\geq 0$ for
all $F_{n}\in \mathcal{F}_{\Lambda _{n}},$ $n=1,2.$

In \emph{the} \emph{quasi classical probability model}, the relation between
a simulation measure $\nu $\ and random variables modelling observables is
such that probabilities of the observed events are reproduced only via
positive values of a normalized bounded real-valued measure $\nu $.

The local version of \emph{the} \emph{quasi classical probability model}
covers (theorem 1) the probabilistic description of each nonsignaling
multipartite correlation scenario, in particular, every multipartite
correlation scenario (corollary 1) on an $N$-partite quantum state.

\section{Appendix A}

\emph{Proof of proposition 1.} Let a scenario $\mathcal{E}_{S}\mathrm{\ }$%
admit an LqHV \ model (\ref{3}). Introduce the normalized real-valued
measure 
\begin{align}
& \mu _{\mathcal{E}_{S}}\left( \mathrm{d}\lambda _{1}^{(1)}\times \cdots
\times \mathrm{d}\lambda _{1}^{(S_{1})}\times \cdots \times \mathrm{d}%
\lambda _{N}^{(1)}\times \cdots \times \mathrm{d}\lambda
_{N}^{(S_{N})}\right)  \tag{A1}  \label{A1} \\
& :=\dint\limits_{\Omega }\text{ }\{\dprod\limits_{s_{n}=1,...,S_{n},\text{ }%
n=1,...,N}P_{n}^{(s_{n})}(\mathrm{d}\lambda _{n}^{(s_{n})}\mid \omega )\text{
}\}\text{ }\nu _{\mathcal{E}_{S}}(\mathrm{d}\omega ).  \notag
\end{align}%
This measure is bounded (see the proof of theorem 1 in \cite{10}) and
returns all distributions $P_{(s_{1},...,s_{N})}^{(\mathcal{E}_{S})}$ of
scenario $\mathcal{E}_{S}$ as the corresponding marginals. The latter means
the factorizable representation%
\begin{align}
P_{(s_{1},...,s_{N})}^{(\mathcal{E}_{S}\mathcal{)}}(F_{1}\times \cdots
\times F_{N})& =\dint \chi _{F_{1}}(\lambda _{1}^{(s_{1})})\cdot ...\cdot
\chi _{F_{N}}(\lambda _{N}^{(s_{N})})\text{ }\mu _{\mathcal{E}_{S}}(\mathrm{d%
}\lambda _{1}^{(1)}  \tag{A2}  \label{A2} \\
& \times \cdots \times \mathrm{d}\lambda _{1}^{(S_{1})}\times \cdots \times 
\mathrm{d}\lambda _{N}^{(1)}\times \cdots \times \mathrm{d}\lambda
_{N}^{(S_{N})}),  \notag \\
F_{1}& \in \mathcal{F}_{\Lambda _{1}},...,F_{N}\in \mathcal{F}_{\Lambda
_{N}},  \notag
\end{align}%
for all $s_{n}=1,...,S_{n},$ $n=1,...,N$. Denote 
\begin{align}
\widetilde{\omega }& :=(\lambda _{1}^{(1)},...,\lambda
_{1}^{(S_{1})},...,\lambda _{N}^{(1)},...,\lambda _{N}^{(S_{N})}),  \tag{A3}
\label{A3} \\
\widetilde{\Omega }& :=\Lambda _{1}^{S_{1}}\times \cdots \times \Lambda
_{N}^{S_{N}},\text{\ \ \ \ }\mathcal{F}_{\widetilde{\Omega }}:=\mathcal{F}%
_{\Lambda _{1}}^{\otimes S_{1}}\otimes \cdots \otimes \mathcal{F}_{\Lambda
_{N}}^{\otimes S_{N}},  \notag \\
\widetilde{\nu }_{\mathcal{E}_{S}}(\mathrm{d}\widetilde{\omega })& :=\mu _{%
\mathcal{E}_{S}}(\mathrm{d}\lambda _{1}^{(1)}\times \cdots \times \mathrm{d}%
\lambda _{1}^{(S_{1})}\times \cdots \times \mathrm{d}\lambda
_{N}^{(1)}\times \cdots \times \mathrm{d}\lambda _{N}^{(S_{N})})  \notag
\end{align}%
and introduce the $\mathcal{F}_{\widetilde{\Omega }}/\mathcal{F}_{\Lambda
_{n}}$-measurable functions $f_{n,s_{n}}:\widetilde{\Omega }\rightarrow
\Lambda _{n},$ each defined by the relation $f_{n,s_{n}}(\widetilde{\omega }%
)=\lambda _{n}^{(s_{n})}.$ Then 
\begin{equation}
\chi _{_{F_{n}}}(\lambda _{n}^{(s_{n})})\equiv \chi
_{f_{n,s_{n}}^{-1}(F_{n})}(\widetilde{\omega }),\text{ \ \ }\forall F_{n}\in 
\mathcal{F}_{\Lambda _{n}},  \tag{A4}  \label{A4}
\end{equation}%
and, in view of (\ref{A3}), (\ref{A4}), representation (\ref{A2}) takes the
form%
\begin{align}
P_{(s_{1},...,s_{N})}^{(\mathcal{E}_{S}\mathcal{)}}(F_{1}\times \cdots
\times F_{N})& =\dint\limits_{\Omega }\chi _{f_{1,s_{1}}^{-1}(F_{1})}(%
\widetilde{\omega })\cdot ...\cdot \chi _{f_{N,s_{_{N}}}^{-1}(F_{N})}(%
\widetilde{\omega })\text{ }\widetilde{\nu }_{\mathcal{E}_{S}}(\mathrm{d}%
\widetilde{\omega })  \tag{A5}  \label{A5} \\
& =\widetilde{\nu }_{\mathcal{E}_{S}}\left( f_{1,s_{1}}^{-1}(F_{1})\cap
\cdots \cap f_{N,s_{_{N}}}^{-1}(F_{N})\right) .  \notag
\end{align}%
This proves the statement of proposition 1.

\section{Appendix B}

\emph{Proof of theorem 1.} If an $N$-partite correlation scenario $\mathcal{E%
}_{S}$, with a setting $S=S_{1}\times \cdots \times S_{N}$, admits a
deterministic LqHV model (\ref{4}), then, clearly, the consistency condition
(\ref{11}) is fulfilled.

Conversely, let scenario $\mathcal{E}_{S}$ satisfy the consistency condition
(\ref{11}). Consider first a bipartite $(N=2)$ scenario $\mathcal{E}_{S}$
with a setting $S=S_{1}\times S_{2}$ and joint probability distributions $%
P_{(s_{1},s_{2})}^{(\mathcal{E}_{S})}$ satisfying condition (\ref{11}).
Since, under condition (\ref{11}), marginals $P_{(s_{1},1)}^{(\mathcal{E}%
_{S})}(F_{1}\times \Lambda _{2}),...,$ $P_{(s_{1},S_{2})}^{(\mathcal{E}%
_{S})}(F_{1}\times \Lambda _{2})$ at site $"1"$ coincide for all $%
s_{2}=1,...,S_{2}$, for simplicity of notation, we denote these coinciding
marginals as%
\begin{equation}
P_{(s_{1},1)}^{(\mathcal{E}_{S})}(F_{1}\times \Lambda
_{2})=...=P_{(s_{1},S_{2})}^{(\mathcal{E}_{S})}(F_{1}\times \Lambda
_{2}):=P_{s_{1}}^{(\mathcal{E}_{S})}(F_{1}),  \tag{B1}  \label{B1}
\end{equation}%
for all $F_{1}\in \mathcal{F}_{\Lambda _{1}}$ and each $s_{1}=1,...,S_{1}$
at site $"1".$ The superscript $\mathcal{E}_{S}$ at notation $P_{s_{1}}^{(%
\mathcal{E}_{S})}$ indicates that, for a general correlation scenario, this
marginal does not need to depend only on a setting of measurement $s_{1}$ at
site $"1"$ (see remark 2).

Quite similarly, 
\begin{equation}
P_{(1,s_{2})}^{(\mathcal{E}_{S})}(\Lambda _{1}\times
F_{2})=...=P_{(S_{1},s_{2})}^{(\mathcal{E}_{S})}(\Lambda _{1}\times
F_{2}):=P_{s_{2}}^{(\mathcal{E}_{S})}(F_{2}),  \tag{B2}  \label{B2}
\end{equation}%
for all $F_{2}\in \mathcal{F}_{\Lambda _{2}}$ and each $s_{2}=1,...,S_{2}$
at site $"2".$

Introduce the normalized bounded real-valued bounded measure $\mu _{\mathcal{%
E}_{S}}$ on $(\Lambda _{1}^{S_{1}}\times \Lambda _{2}^{S_{2}},$ $\mathcal{F}%
_{\Lambda _{1}}^{\otimes S_{1}}\otimes \mathcal{F}_{\Lambda _{2}}^{\otimes
S_{2}})$ with values%
\begin{align}
& \mu _{\mathcal{E}_{S}}\text{ }(F_{1}^{(1)}\times \cdots \times
F_{1}^{(S_{1})}\times F_{2}^{(1)}\times \cdots \times F_{2}^{(S_{2})}) 
\tag{B3}  \label{B3} \\
& :=\sum_{s_{1},s_{2}}\left\{ P_{(s_{1},s_{2})}^{(\mathcal{E}%
_{S})}(F_{1}^{(s_{1})}\times F_{2}^{(s_{2})})\tprod\limits_{\widetilde{s}%
_{1}\neq s_{1}}P_{\widetilde{s}_{1}}^{(\mathcal{E}_{S})}(F_{1}^{(\widetilde{s%
}_{1})})\tprod\limits_{\widetilde{s}_{2}\neq s_{2}}P_{\widetilde{s}_{2}}^{(%
\mathcal{E}_{S})}(F_{2}^{(\widetilde{s}_{2})})\right\}  \notag \\
& -\text{ }(S_{1}S_{2}-1)\tprod\limits_{s_{1}}P_{s_{1}}^{(\mathcal{E}%
_{S})}(F_{1}^{(s_{1})})\tprod\limits_{s_{2}}P_{s_{2}}^{(\mathcal{E}%
_{S})}(F_{2}^{(s_{2})}),  \notag
\end{align}%
for all $F_{n}^{(s_{n})}\in \mathcal{F}_{\Lambda _{n}},$ $s_{n}=1,...,S_{n},$
$n=1,2.$ It is easy to verify that this measure returns all joint
probability distributions $P_{(s_{1},s_{2})}^{(\mathcal{E}_{S})}$ of a
bipartite nonsignaling scenario $\mathcal{E}_{S}$ as the corresponding
marginals. By proposition 2, this implies that a bipartite correlation
scenario satisfying condition (\ref{11}) admits a deterministic LqHV model.

Let $N=3.$ Consider a tripartite correlation scenario $\mathcal{E}_{S}$ with
a setting $S=S_{1}\times S_{2}\times S$ and joint probability distributions $%
P_{(s_{1},s_{2},s_{_{N}})}^{(\mathcal{E})}$ satisfying condition (\ref{11}).
In addition to the one-party marginals denoted similarly to notation (\ref%
{B2}), we denote by $P_{(s_{1},s_{2})}^{(\mathcal{E}_{S})},$ $%
P_{(s_{1},s_{3})}^{(\mathcal{E}_{S})},$ $P_{(s_{2},s_{3})}^{(\mathcal{E})}$
the coinciding two-party marginals at the corresponding sites, that is:%
\begin{align}
P_{(s_{1},s_{2},1)}^{(\mathcal{E}_{S})}(F_{1}\times F_{2}\times \Lambda
_{3})& =...=P_{(s_{1},s_{2},S_{3})}^{(\mathcal{E}_{S})}(F_{1}\times
F_{2}\times \Lambda _{3}):=P_{(s_{1},s_{2})}^{(\mathcal{E}_{S})}(F_{1}\times
F_{2}),  \tag{B4}  \label{B4} \\
P_{(s_{1},1,s_{3})}^{(\mathcal{E}_{S})}(F_{1}\times \Lambda _{2}\times
F_{3})& =...=P_{(s_{1},S_{2},s_{3})}^{(\mathcal{E}_{S})}(F_{1}\times \Lambda
_{2}\times F_{3}):=P_{(s_{1},s_{3})}^{(\mathcal{E}_{S})}(F_{1}\times F_{3}),
\notag \\
P_{(1,s_{2},s_{3})}^{(\mathcal{E}_{S})}(\Lambda _{1}\times F_{2}\times
F_{3})& =...=P_{(S_{1},s_{2},s_{3})}^{(\mathcal{E}_{S})}(\Lambda _{1}\times
F_{2}\times F_{3}):=P_{(s_{2},s_{3})}^{(\mathcal{E})}(F_{2}\times F_{3}), 
\notag
\end{align}%
for all $F_{n}\in \mathcal{F}_{\Lambda _{n}},$ $s_{n}=1,...,S_{n},$ $%
n=1,2,3. $

Similarly to our construction of measure (\ref{B3}) for a bipartite case,
introduce the normalized bounded real-valued measure $\widetilde{\mu }_{%
\mathcal{E}_{S}}$ on $(\Lambda _{1}^{S_{1}}\times \Lambda _{2}^{S_{2}}\times
\Lambda _{3}^{S_{3}},$ $\mathcal{F}_{\Lambda _{1}}^{\otimes S_{1}}\otimes 
\mathcal{F}_{\Lambda _{2}}^{\otimes S_{2}}\otimes \mathcal{F}_{\Lambda
_{3}}^{\otimes S_{3}})$ with values

\begin{align}
& \widetilde{\mu }_{\mathcal{E}_{S}}\text{ }(F_{1}^{(1)}\times \cdots \times
F_{1}^{(S_{1})}\times F_{2}^{(1)}\times \cdots \times F_{2}^{(S_{2})}\times
F_{3}^{(1)}\times \cdots \times F_{3}^{(S_{3})})  \tag{B5}  \label{B5} \\
&  \notag \\
& :=\sum_{s_{1},s_{2},s_{_{3}}}\left\{ P_{(s_{1},s_{2},s_{_{3}})}^{(\mathcal{%
E}_{S})}(F_{1}^{(s_{1})}\times F_{2}^{(s_{2})}\times
F_{3}^{(s_{3})})\tprod\limits_{\widetilde{s}_{1}\neq s_{1}}P_{\widetilde{%
s_{1}}}^{(\mathcal{E}_{S})}(F_{1}^{(\widetilde{s}_{1})})\tprod\limits_{%
\widetilde{s}_{2}\neq s_{2}}P_{\widetilde{s}_{2}}^{(\mathcal{E}%
_{S})}(F_{2}^{(\widetilde{s}_{2})})\tprod\limits_{\widetilde{s}_{3}\neq
s_{3}}P_{\widetilde{s}_{3}}^{(\mathcal{E}_{S})}(F_{3}^{(\widetilde{s}%
_{3})})\right\}  \notag \\
&  \notag \\
& -(S_{1}-1)\tprod\limits_{s_{1}}P_{s_{1}}^{(\mathcal{E}%
_{S})}(F_{1}^{(s_{1})})\sum_{s_{2},s_{3}}\text{ }\left\{ P_{(s_{2},s_{3})}^{(%
\mathcal{E}_{S})}(F_{2}^{(s_{2})}\times F_{3}^{(s_{3})})\tprod\limits_{%
\widetilde{s}_{2}\neq s_{2}}P_{\widetilde{s}_{2}}^{(\mathcal{E}%
_{S})}(F_{2}^{(\widetilde{s}_{2})})\tprod\limits_{\widetilde{s}_{3}\neq
s_{3}}P_{\widetilde{s}_{3}}^{(\mathcal{E}_{S})}(F_{3}^{(\widetilde{s}%
_{3})})\right\}  \notag \\
&  \notag \\
& -(S_{2}-1)\tprod\limits_{s_{2}}P_{s_{2}}^{(\mathcal{E}%
_{S})}(F_{2}^{(s_{2})})\sum_{s_{1},s_{3}}\left\{ P_{(s_{1},s_{3})}^{(%
\mathcal{E}_{S})}(F_{1}^{(s_{1})}\times F_{3}^{(s_{3})})\tprod\limits_{%
\widetilde{s}_{1}\neq s_{1}}P_{\widetilde{s}_{1}}^{(\mathcal{E}%
_{S})}(F_{1}^{(\widetilde{s}_{1})})\tprod\limits_{\widetilde{s}_{3}\neq
s_{3}}P_{\widetilde{s}_{3}}^{(\mathcal{E}_{S})}(F_{3}^{(\widetilde{s}%
_{3})})\right\}  \notag \\
&  \notag \\
& -(S_{3}-1)\tprod\limits_{s_{3}}P_{s_{3}}^{(\mathcal{E}%
_{S})}(F_{3}^{(s_{3})})\sum_{s_{1},s_{2}}\left\{ P_{(s_{1},s_{2})}^{(%
\mathcal{E}_{S})}(F_{1}^{(s_{1})}\times F_{2}^{(s_{2})})\tprod\limits_{%
\widetilde{s}_{1}\neq s_{1}}P_{\widetilde{s}_{1}}^{(\mathcal{E}%
_{S})}(F_{1}^{(\widetilde{s}_{1})})\tprod\limits_{\widetilde{s}_{2}\neq
s_{2}}P_{\widetilde{s}_{2}}^{(\mathcal{E}_{S})}(F_{2}^{(\widetilde{s}%
_{2})})\right\}  \notag \\
&  \notag \\
&
+(2S_{1}S_{2}S_{3}-S_{1}S_{2}-S_{2}S_{3}-S_{1}S_{3}+1)\tprod%
\limits_{s_{1}}P_{s_{1}}^{(\mathcal{E}_{S})}(F_{1}^{(s_{1})})\tprod%
\limits_{s_{2}}P_{s_{2}}^{(\mathcal{E}_{S})}(F_{2}^{(s_{2})})\tprod%
\limits_{s_{3}}P_{s_{3}}^{(\mathcal{E}_{S})}(F_{3}^{(s_{3})}),  \notag \\
&  \notag
\end{align}%
for all sets $F_{n}^{(s_{n})}\in \mathcal{F}_{\Lambda _{n}},$ $%
s_{n}=1,...,S_{n},$ $n=1,2,3.$ Measure $\widetilde{\mu }_{\mathcal{E}_{S}}$
returns all joint probability distributions $P_{(s_{1},s_{2},s_{_{3}})}^{(%
\mathcal{E}_{S})}$ of a \ tripartite nonsignaling scenario $\mathcal{E}_{S}$
as the corresponding marginals. By proposition 2, the latter implies that a
correlation scenario $\mathcal{E}_{S}$ satisfying condition (\ref{11})
admits a deterministic LqHV model.

The obvious generalization to an arbitrary $N$-partite case of the measure
constructions used in (\ref{B3}), (\ref{B5}) proves the sufficiency part of
theorem 1.

\end{document}